\documentclass[10pt,journal,compsoc]{IEEEtran}

\ifCLASSOPTIONcompsoc
  \usepackage[nocompress]{cite}
\else
  \usepackage{cite}
\fi

\usepackage{mathptmx}
\usepackage{amsmath}
\usepackage{amssymb}
\usepackage{amsfonts}
\usepackage{graphicx}
\usepackage{times}
\usepackage{pifont}
\usepackage{comment}
\usepackage{algorithm}
\usepackage{algpseudocode}
\usepackage{gensymb}
\usepackage{wrapfig}
\usepackage{array}
\usepackage{soul}
\usepackage{cancel}
\usepackage{color}
\usepackage{slashbox}
\usepackage{multirow}
\usepackage{array}
\usepackage{hyperref}
\usepackage{CJKutf8}
\usepackage[normalem]{ulem}

\usepackage{textcomp}
\usepackage{dblfloatfix}
\usepackage{booktabs}
\usepackage{color}
\usepackage{ dsfont }
\usepackage{ulem}
\usepackage{listings}
\usepackage{adjustbox}
\usepackage{enumitem}

\newcommand{\PreserveBackslash}[1]{\let\temp=\\#1\let\\=\temp}
\newcolumntype{C}[1]{>{\PreserveBackslash\centering}p{#1}}
\newcolumntype{R}[1]{>{\PreserveBackslash\raggedleft}p{#1}}
\newcolumntype{L}[1]{>{\PreserveBackslash\raggedright}p{#1}}

\usepackage{pifont}%

\usepackage{calc}
\newlength\myheight
\newlength\mydepth
\settototalheight\myheight{Xygp}
\usepackage[normalem]{ulem}

\newcommand{\revision}[1]{\leavevmode{\color{blue}{#1}}}
\newcommand{\remark}[1]{\textcolor[RGB]{150, 200, 0}{#1}}

\setstcolor{red}
\newcommand{\removed}[1]{\leavevmode{\color{red}{\st{#1}}}}

\def \cleanversion{} %
\ifx\cleanversion\undefined
\else
  \renewcommand{\remark}[1]{} %
  \renewcommand{\removed}[1]{} 
  \renewcommand{\revision}[1]{#1}
  
\fi

\begin{document}

\begin{CJK}{UTF8}{gbsn}

\title{AutoLegend: A User Feedback-Driven Adaptive Legend Generator for Visualizations}

\author{Can~Liu,
        Xiyao~Mei,
        Zhibang~Jiang,
        Shaocong~Tan,
        Xiaoru~Yuan%
\IEEEcompsocitemizethanks{
\IEEEcompsocthanksitem Can Liu, Xiyao Mei, Shaocong Tan, and Xiaoru Yuan are with Key Laboratory of Machine Perception (Ministry of Education), School of AI, Peking University. E-mail: \{can.liu, xiyao.mei, shaocong.tan, xiaoru.yuan\}@pku.edu.cn.
\IEEEcompsocthanksitem Zhibang Jiang is with Parsons School of Design, The New School. E-mail: zhibang.jinag@newschool.edu.}%
\thanks{Manuscript received xx xx, 2023; revised xx xx, 2023.}}

\markboth{TRANSACTIONS,~Vol.~XX, No.~XX, August~2023}%
{Liu et al.: AutoLegend}

\IEEEtitleabstractindextext{%
\begin{abstract}
We propose AutoLegend to generate interactive visualization legends using online learning with user feedback.
AutoLegend accurately extracts symbols and channels from visualizations and then generates quality legends.
AutoLegend enables a two-way interaction between legends and interactions, including highlighting, filtering, data retrieval, and retargeting.
After analyzing visualization legends from IEEE VIS papers over the past 20 years, we summarized the design space and evaluation metrics for legend design in visualizations, particularly charts. The generation process consists of three interrelated components: a legend search agent, a feedback model, and an adversarial loss model. The search agent determines suitable legend solutions by exploring the design space and receives guidance from the feedback model through scalar scores. The feedback model is continuously updated by the adversarial loss model based on user input.
The user study revealed that AutoLegend can learn users' preferences through legend editing.

\end{abstract}

}

\maketitle

\IEEEdisplaynontitleabstractindextext

\section{Introduction}

Legends play an indispensable role in data visualizations, enabling users to grasp the mapping relationship between data attributes and visual channels. They shed light on the purpose and significance of visualizations by illustrating the attributes and data ranges involved. Despite the crucial role of legends in visualization, a significant number of static visualizations lack proper legends or have inaccurately designed ones, a trend observed even in academic papers and widely used tools.
While numerous visualization toolkits offer features for generating legends, instances of inadequate or missing legends remain prevalent. This is attributed to the time and effort required to implement a well-designed legend. The challenges inherent in crafting effective legends stem from the intricate nature of the legend design domain and the lack of standardization.
On one hand, the design landscape for legends is vast, encompassing multiple dimensions such as visual channels, visual marks, element layout, text arrangement, and layout considerations when multiple symbols or channels are involved. The amalgamation of diverse options across these dimensions creates an extensive design realm, complicating the identification of the optimal design solution.
On the other hand, the absence of a standardized approach to legend design results in varying preferences among different creators. Furthermore, the dearth of robust evaluation metrics exacerbates the situation. Without well-defined evaluation criteria, the task of comparing and identifying the most effective legend designs becomes challenging.

We propose a method for automatically generating effective interactive legends, taking user feedback into account.
The understanding of the design space is the beginning of an automatic method.
We examined visualizations from the past two decades of IEEE VIS, summarizing the design space of visualization legends.
The visualization legends design space consists of five dimensions, namely, visual marks, visual channels, symbol layout, text layout, and multi-channel layout. 
To identify the most common problems, we surveyed legends from a gallery of widely-used visualization tools (e.g., D3~\cite{bostock2011d3}, Vega-lite~\cite{satyanarayan2017vegalite}, and EChart~\cite{li2018echarts}).
Among the 702 visualizations, there are only 466 with legends;
Among those visualization with legends, 53 are with inappropriate legends.
We also examined student assignments to identify inappropriate legends.
Among the 1,368 visualizations, there are 122 with improper legends.
Moreover, we proposed evaluation metrics for visualization legends, such as preventing overlapping of existing channels, ink balance, space occupation, legend organization, legend text readability, and space correspondence.

Based on the design space and evaluation metrics, we developed a human-AI collaborated visualization legend generator, AutoLegend. 
Taking a visualization as input, AutoLegend generates a legend that adheres to the evaluation metrics as closely as possible.
The legend generation process comprises three parts: extracting iconic symbols and mapping channels, searching for suitable solutions in a high-dimensional legend space, and scoring these solutions based on a reward model.
The reward model takes evaluation metrics as inputs and outputs a comprehensive score through a multi-layer neural network.
The search network, which is based on a genetic algorithm, supports searching for a solution in a mixed space comprising discrete space (e.g., arrangement direction, symbol layout, text layout) and continuous space (e.g., position).
Acknowledging that different users may have distinct preferences, the reward model is also incorporated preference metrics (e.g., horizontal, vertical, center, or edge preference).
Our approach achieves a machine learning method in a human-in-the-loop framework that supports online user interaction adjustment and the quality assessment model's dynamic updating.

In summary, our contributions are as follows:
\begin{itemize}
\item We analyzed the design space of visualization legends and identified five dimensions that encompass the scope of visualization legends.
\item We developed a tool, AutoLegend, a real-time feedback system for generating legends for visualizations by extracting the marks and channels and deciding the position, symbol layouts, text layouts, and multi-legend layouts. AutoLegend enables users to modify the legend according to their preferences and updates the backend model accordingly.
\end{itemize}

In \autoref{sec: back}, we discuss the related work of AutoLegend. \autoref{sec:design_space} summarizes and introduces the design space for visualization legends. In \autoref{sec: assess}, we present the evaluation metrics for assessing visualization legends. Our automatic visualization legend generation method, based on the design space and evaluation metrics with a human-in-the-loop approach, is detailed in \autoref{sec: generation}. Subsequently, \autoref{sec: legendinter} describes the interactions supported by AutoLegend-generated legends. \autoref{sec: cases} showcases a variety of legend examples generated by AutoLegend. The user study of AutoLegend is outlined in \autoref{sec: study}. In \autoref{sec: discuss}, we discuss potential future directions. We conclude our work in \autoref{sec: conclusion}.

\section{Related Work}
\label{sec: back}

AutoLegend is a system generating interactive legends for visualizations.
As AutoLegend extracts the content of visualization and supports interaction with visualization on the legend, it is also related to the extraction and interaction enhancement.

\subsection{Visualization Legends}

The previous works about visualization legends are mainly based on maps.
As maps rapidly evolve towards multimedia, three-dimensional visualization, and interactivity, Sieber~\cite{sieber2005smart} proposed the concept of a ``smart legend," which serves as a central control unit for digital maps.
This approach encompasses self-acting adaptations for optimal map element depiction and a range of user interaction features.
Gobel et al.~\cite{gobel2018gazelegend} focused on improving interaction with map legends by adapting their placement and content based on the user's gaze.
Dykes et al.~\cite{dykes2010rethinkmaplegend} presented guidelines for legend design in a visualization context derived from cartographic literature. These guidelines address selection, layout, symbols, position, dynamism, design, and process, which can be applied to various legends and keys in cartography and information visualization.
Regarding legend placement, Imhof~\cite{imhof1972thematische} introduced the concept of visual weight and suggested positioning the legend opposite the visual center. Edler et al.~\cite{edler2020searching} found that legends positioned to the right of the map field are decoded faster without impairing recognition memory performance.

In recent years, interactive legends have emerged as an effective means to facilitate user interaction with visualizations.
Riche et al.~\cite{riche2010understanding} demonstrated that interactive legends not only improve the perception of mapping between data values and visual encoding but also affect interaction time differently depending on the data type. Furthermore, their study highlighted the superiority of ordinal controls over numerical techniques, which are predominantly used in today's systems.

\subsection{Visualization Extraction}

The extraction of data from visualizations\cite{shahira2021towards} has been a topic of considerable interest in recent years, with several researchers focusing on developing methods to improve the accessibility\cite{Vis4nonvisual}, searchability, and reusability of visualizations.
Savva et al.~\cite{savva2011revision} and Poco et al.~\cite{poco2017reverse} proposed methods for reverse-engineering visualizations from bitmap images, which identify the classification of charts and the positions of visual elements to extract the mapping relationships of visualizations by analyzing the information in images.
However, these methods mainly focus on the information of axis mapping and lack the analysis of the information of the legend.

To address these limitations, more recent works have utilized deep learning methods for information extraction from visualizations.
For example, Yuan et al.~\cite{yuan2022colormap} used deep learning methods to extract color mapping, while Luo et al.~\cite{Luo2021WACV} used deep learning methods to extract textual and graphical information from charts.
Similarly, Lai et al.~\cite{Lai2020Annotation} employed OCR methods to extract visual elements from visualization charts, while Zhou et al.~\cite{zhou2021reverse} used neural networks to extract information from bar charts. Zhang et al.~\cite{Zhang2020MI3} focused on extracting data from ancient visualizations with greater diversity, using a combination of interactive and machine learning methods for data extraction.
Liu et al.~\cite{liu2019data} used a single neural network for information extraction from visualizations.
Poco et al.~\cite{Poco2018extract} proposed a method for extracting color mappings from bitmap images.
Hoque et al.~\cite{hoque2019searching} supported the reuse of D3 visualizations. 
Cui et al.~\cite{cui2021mixed} extracted reusable templates from information graphics, while Chen et al.~\cite{Chen2020timeline} extracted corresponding templates from timelines.
These efforts aim to enhance the usability and accessibility of visualizations, empowering users to create, restyle, or interact with visualization with less effort.

\subsection{Interaction Enhancement}

Interaction enhancement approaches for visualization have been proposed for adding animations to the existing visualizations to increase the readability or emphasize specific data attributes.
Kong and Agrawala~\cite{kong2012graphical} added kinds of animations to visualizations to improve readability.
Lu et al.~\cite{Lu2021ant} emphasizes data attributes on static charts by encoding data attributes with animations.

In recent years, some approaches have aimed to enhance the existing visualizations through interactions or animations.
VisDock\mbox{\cite{choi2015visdock}} was proposed as a system that allows the programmer to add interactions (e.g., selecting, filtering, navigation, etc.) on existing visualizations with codes.
Based on features of D3 specifications on DOM elements, Harper and Agrawala\mbox{\cite{harper2014deconstructing, harper2018converting}} developed tools deconstructing existing D3 visualizations by matching the given data with visual attributes.
The extracted mapping relationship enables reusing with templates of Vega-lite\mbox{\cite{satyanarayan2017vegalite}}. 
Interaction+\mbox{\cite{lu2017interaction}} enhanced interactions for visualization on the web by parsing the attributes of visual marks and applying extra interaction add-ons to the visualization.
Interaction+ focuses on non-spatial attributes like color and opacity, while Liu et al.~\cite{liu2024spatial} proposed a spatial-constraint-based method for adding spatially-related interactions to static visualizations.

\begin{figure}[!ht]
  \centering
  \includegraphics[width=\columnwidth]{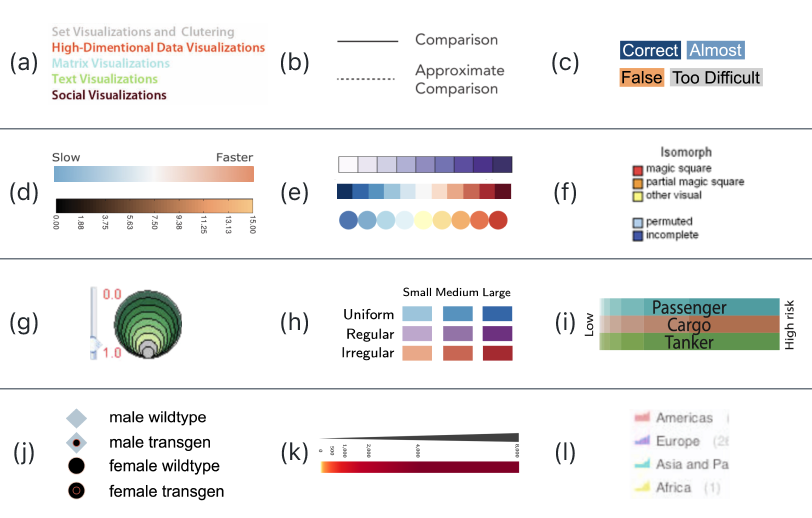}
  \caption{
Representative legends from IEEE VIS papers.
(a) Text as symbol \cite{goodcase_a};
(b) Text accompanying a symbol \cite{goodcase_b};
(c) Text-embedded symbol\cite{goodcase_c};
(d) Continuous color legend \cite{goodcase_d_up, goodcase_d_down};
(e) Connected symbol layout \cite{goodcase_e_up, goodcase_e_middle_down};
(f) Non-uniform symbol layout  \cite{goodcase_f};
(g) Nested symbol layout\cite{goodcase_g};
(h) Matrix form with color-discrete symbol \cite{goodcase_h};
(i) Continuous matrix form \cite{goodcase_i};
(j) Flattened form \cite{goodcase_j};
(k) Paralleled form \cite{goodcase_k};
(l) Data-encoded symbol\cite{goodcase_l}.
 }
  \label{fig: goodlegend}
\end{figure}

\section{Design Space of Legends}
\label{sec:design_space}

\begin{figure*}[!ht]
  \centering
  \includegraphics[width=\textwidth]{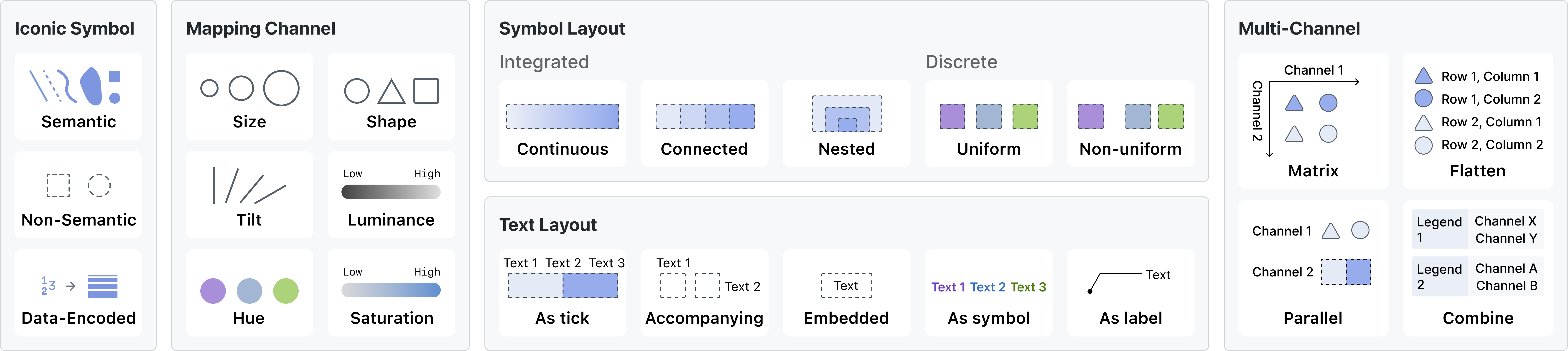}
  \caption{The design space of a legend refers to the multi-dimensional space constituted by different design options of a legend. It includes five dimensions: the iconic symbol type of the legend, the mapping channel, the symbol layout, the text layout, and the layout of multiple channel legends.}
  \label{fig: space}
\end{figure*}

To investigate the design space of legend, we reviewed 12,267 images on 1,397 VAST and InfoVis publications in the past 22 years from the VisImages~\cite{deng2022visimages} dataset.
Of these images, only 2,327 ($\sim $ 19\%) are with legends, while the rest are lost for various reasons. 
The authors may have overlooked the necessity of legends or not invested the effort to create them owing to time constraints.
This shows the importance of automatic legend generation for visualization stakeholders. We identified 2,392 legends from the images and categorized them according to their visual encoding, spatial layout, and text position. \autoref{fig: goodlegend} shows some examples of the captured legends. Subsequently, we summarized the design space shown in \autoref{fig: space} and labeled the five dimensions among these legends. The annotations are available on the website\footnote{Legends in VIS Publications: \url{https://autolegend.github.io/legends_in_vis_publications/index.html}}.

To investigate the use of legends in visualization practice, we collected 1,368 student projects from two visualization courses over the past 10 years (2012-2022), as well as visualizations from three galleries of common libraries, including 168 from the D3 gallery\footnote{D3 gallery: \url{https://observablehq.com/@d3/gallery}}, 339 from Echarts\footnote{Echarts gallery: \url{https://echarts.apache.org/examples/}}, and 195 from Vega-lite\footnote{Vega-lite gallery: \url{https://vega.github.io/vega-lite/examples/}}.
Out of the 2,070 visualizations we gathered, 1,256 lacked accompanying legends. Among the visualizations that did have legends, 175 exhibited inconsistencies. The inconsistency here pertains to situations where the legend and the chart display visual encodings that diverge, undermining the user's capacity to establish visual correlations and, in some cases, resulting in the misinterpretation of data. Such inconsistency issues predominantly manifest in the Iconic Symbol and Mapping Channel dimensions. The set of visualizations featuring these problematic legends has been made available to the public\footnote{Inconsistent Legend: \url{https://autolegend.github.io/inconsistent_legends/}}.
Creating appropriate legends in the proper scenarios remains a challenging task for both visualization novices and experts. 
Common issues we observed in the legends included unclear or misleading labels, confusing symbols or colors, and inconsistent use of legends across multiple visualizations. These findings highlight the need for better education and guidance on legend design in visualization courses and for the development of automated approaches to assist users in generating appropriate and consistent legends.

\subsection{Iconic symbol and Mapping channel}

A visualization legend includes iconic symbols and channels used for mapping. Iconic symbols can be categorized into semantic, non-semantic, and data-encoded. Semantic symbols use similar shapes to represent visual elements, while non-semantic symbols provide no shape correspondence. Data-encoded symbols add an extra dimension to present additional data attributes.
Mapping channels include size, position, rotation, and color. The legend contains two types of channels: those used for mapping data attributes and those that remain constant. Among the 2,392 legends on VIS papers, there are 886 with \textit{semantic} symbol, 1,505 with \textit{non-semantic} symbol, and only 1 with \textit{data-encoded} symbol.

\subsection{Symbol Layout}

We summarize four distinct types of visual representations: continuous, connected, nested, and discrete.

\begin{itemize}

\item \textbf{Continuous} symbols manifest as elongated forms utilizing continuous visual channels, adept at representing quantitative data attributes.
Widely employed methods encompass leveraging brightness, saturation, and other continuous trajectories within color space. Within continuous distributions, various monotonicity alternatives are available, including unidirectional monotonic sequences, bi-directional sequences anchored by a discernible zero point, and distinctive shifts in both directions emanating from the zero point—illustrated, for instance, by distinct hues denoting regions above and below sea level.

\item \textbf{Connected} layout comprises multiple symbols linked together in an elongated shape, maintaining a consistent visual channel. It is frequently used for quantitative data attributes and can represent discrete quantitative data or sampled continuous quantitative data. Connected graphics typically have equal length, although some might use length to represent statistical values or embedded text length.

\item \textbf{Nested} symbols feature multiple nested, stacked sampled elements arranged with larger elements below and smaller ones above. To avoid complete coverage by lower elements, nested representations are generally used to depict size-related visual channels.

\item \textbf{Discrete} layout consists of multiple sampled symbols organized in a list, with each text corresponding to a symbol.
It is primarily used for categorical and ordinal data attributes.
The symbols are usually evenly spaced, but in some cases, they may be unevenly spaced to correspond to visual elements. Line breaks may occur when there is a constraint on the space position.
\end{itemize}

Among the 2,392 legends on VIS papers, there are 464 with \textit{continuous} symbol, 422 with \textit{connected} symbol, 1 with \textit{nested} symbol, 1,505 with \textit{discrete} symbol (1,351 \textit{uniform} and 154 \textit{non-uniform}).

\subsection{Text Layout}

The text layout within a visualization legend serves diverse roles, such as serving as ticks, accompanying graphic elements, overlaying graphic elements, representing the color as a graphic element, and freely spaced serving as labels.

\begin{itemize}
    \item \textbf{As tick}: Texts can play the role of ticks on the legend. Tick-text presents the value of the sampling position of the legend. It represents the value of a sampling position or the boundary. The value of other positions in the legend requires the users to interpolate. Tick-text is suitable for continuous or connected legends encoding the quantitative attributes.
    \item \textbf{Accompanying}: Texts can be displayed alongside their corresponding symbols, representing the symbol's value. There are two options for the arrangement of the text: cross-arrangement and side-by-side arrangement. In cross-arrangement, the text and symbols appear in the same row or column and cross over each other. On the other hand, in a side-by-side arrangement, the text is arranged in a separate sequence parallel to the symbol list. Cross-layout makes the legend thinner, making it more suitable for placement at the edges of the visualization. At the same time, the side-by-side arrangement can maintain the correspondence between symbols and texts.
    
    \item \textbf{Embedded}: Texts may overlay on top of their corresponding symbols, indicating the semantics of the symbols. The overlay textual information presents a high correspondence between the text and the symbol. However, when the color of the text and symbol is similar, legibility may be reduced.

    \item \textbf{As symbol}: The textual component itself can serve as a symbolic representation, wherein the text is endowed with a color matching that of the visual elements it correlates with. This correspondence is conveyed through the alignment of text color with that of the associated visual elements. However, this alignment, characterized by the proximity between font color and background color, may potentially undermine legibility.

    \item \textbf{As label}: Text can also express correspondence through auxiliary lines or positional relations, with this type of organization employed using existing visual elements as symbols. The label-like legends are not the common form of legends and can be considered annotations or labels.
\end{itemize}

Of the 2,392 legends that appeared on VIS publications, 592 had texts \textit{as tick}, 1,467 had \textit{accompanying} texts, 133 had the \textit{embedded} texts, 63 had texts \textit{as symbol}, 1 had texts \textit{as label}, and others had no text within legends.

\subsection{Multi-Legend Layout}

When a visualization legend contains multiple symbols or a single symbol has multiple channels, a single legend may not be sufficient to map the relationships between them.
To address this issue, a matrix can be used to represent these relationships by treating the multiple channels of a single symbol as different dimensions of the matrix.
When the number of possibilities is small, the matrix form or an expanded form may be adopted. However, when the number is large, different dimensions of the legend can be treated as separate legends and placed parallelly.

\begin{itemize}
\item \textbf{Matrix}: A matrix-shaped legend is constructed through two vertically arranged visual channels of the same mark.
This form usually has a small number of discrete or continuous values and traverses all possible value options.
\item \textbf{Flattened}: A flattened representation of a matrix form is also a way to express multiple channels.
This form resembles a typical list form but encodes information across two or more dimensions. \remark{Flatten ba be confused with matrix, depending on the semantic.}
\item \textbf{Parallel}: \remark{have a difficulty understanding the parallel layout.} The two channels are expressed separately and placed in parallel.
This approach decouples the two channels of the same visual element. Users need to interpret them separately.
\item \textbf{Combined}: This structure places two legends parallel to each other, which may correspond to different visual elements in the visualization, such as the color of the area and the color of the edges. A clear indication is necessary to differentiate the different visual elements represented by each legend.
\end{itemize}

Among the 390 multi-channel legends in 2,392 legends of VIS papers, 50 had texts \textit{matrix} layout, 74 were \textit{flattened} layout, 47 were \textit{parallel} layout, and 219 were \textit{combined} layout.

\section{Assessment of Legend Quality}
\label{sec: assess}

In this section, we summarize the evaluation metrics for visualization legends based on common issues that arise in visualizations. These metrics include the principles of non-obstruction of visual elements, visual balance, text legibility, and correspondence.
\remark{The paper would be strengthened if the authors can provide additional supporting arguments why those metrics reflect the judgement of the human reader.}

\textbf{Obstruction Reduction}: An effective legend should not hinder the presentation of crucial information. To prevent the obstruction of critical details, we compute the standard deviation of pixel values in the region where the legend is situated, quantifying the legend's coverage degree. A standard deviation of 0 indicates a homogeneous region. When assessing the homogeneity of the legend's region, we calculate the mean pixel value of the R, G, and B channels within the legend's region in the original image, denoted by $\mu$. The width and height of the legend are represented by $w$ and $h$, respectively, while the pixel value at coordinate ($i$, $j$) of the corresponding channel in the legend is denoted by $p_{ij}$. The degree of obstruction can be defined as:
$$\mu = \frac{1}{w\times h}\sum_{i}^{w}\sum_{j}^{h}p_{ij},$$
$$O = \sqrt[]{\frac{1}{w\times h}\sum_{i}^{w}\sum_{j}^{h}(p_{ij}-\mu )^2 }.$$

\textbf{Ink Balance}: The integration of a legend should enhance the ink distribution in a visualization. Imhof~\cite{imhof1972thematische} introduced the concept of visual weight, suggesting the placement of the legend opposite the map's visual weight center. The overall balance of the visualization should be improved with the inclusion of a legend. Spatial balance is often associated with the centroid of ink weight. The closer the centroid of ink weight is to the geometric center, the more balanced the image. To compute the ink-weight centroid, the RGB image is converted to a grayscale image, with the white color regarded as zero. \revision{The visual weight centroid is calculated based on the grayscale image, where the gray value at coordinate ($i$, $j$) is $g_{ij}$.} The ink-weight balance of the image is measured by the distance between the centroid of the ink-weight and the geometric center. The ink balance metric $I$ is defined as: 
$$x' = \frac{\sum_{i=1}^{w}(i\sum_{j=1}^{h}g_{ij})}{\sum_{i=1}^{w}\sum_{j=1}^{h}g_{ij}},\\$$
$$y' = \frac{\sum_{j=1}^{h}(j\sum_{i=1}^{w}g_{ij})}{\sum_{i=1}^{w}\sum_{j=1}^{h}g_{ij}},$$
$$I = \sqrt[]{(x'-x_c)^2+(y'-y_c)^2}.$$

\textbf{Text Readability}: 
\revision{Texts are usually used to show visual channel of the legend. In the design space of legend, texts can overlay or accompany the legends, even serve as legends. Thus, the readability of the texts of legends should be of high quality, avoiding any font identification issues that may result from mismatched foreground and background colors. Here, text readability is measured by the contrast ratio $R$ between the text and its background. W3C\footnote{W3C Web Content Accessibility Guidelines (WCAG) 2.0: \url{https://www.w3.org/TR/WCAG20/}\label{w3c}} defines contrast ratio $R$ as follows, where $L_{min}$ is the relative luminance of the lighter of the colors and $L_{min}$ is the relative luminance of the darker of the colors. The formula to calculate relative luninance can be found in W3C\textsuperscript{\ref{w3c}} as well.
$$R = \frac{L_{max} + 0.05}{L_{min} + 0.05}.$$ 
The contrast ratio typically ranges from 1 to 21, according to W3C\textsuperscript{\ref{w3c}}, it should be at least 4.5:1 to achieve a better readability.
}

\textbf{Size Minimization}:
When including a legend in a visualization, it's important to avoid introducing a significant amount of unused space. For example, adding a wide legend to the right of the visualization can make it wider and create a large amount of unnecessary space. To evaluate the effectiveness of a legend, we calculate the ratio of the bounding box after the legend is added to the original bounding box. Specifically, we define the ``size minimization'' ($S$) as:
$$S = \frac{Area_{vis\&legend} - Area_{vis}}{Area_{vis}}.$$
where the area of the bounding box is computed as the product of its width and height.
A higher value of $S$ indicates a larger increase in unused space due to the addition of the legend, which reduces the effectiveness of the information conveyed by the visualization.

\textbf{Correspondence Principle}: The Correspondence Principle emphasizes the importance of maintaining a clear and consistent relationship between the visual elements in visualization and the symbols in its accompanying legend. This involves ensuring that there is correspondence between color, shape, and spatial location in the two components.
Color correspondence requires using the same colors for the symbols in the legend as those used in the visualization, allowing users to easily identify which parts of the legend correspond to which visual elements in the visualization.
Spatial location correspondence has two components: structural correspondence and proximity correspondence. Structural correspondence ensures that symbols and visual elements have the same order, enabling users to easily find the corresponding legend item for a particular visual element. The Correspondence Principle is defined as the summation of correspondence in color, shape, and symbol order, represented as $$C = C_{color} + C_{shape} + C_{order}.$$

\section{Automatic Legend Generation}
\label{sec: generation}

\begin{figure*}[!ht]
    \centering
    \includegraphics[width=\textwidth]{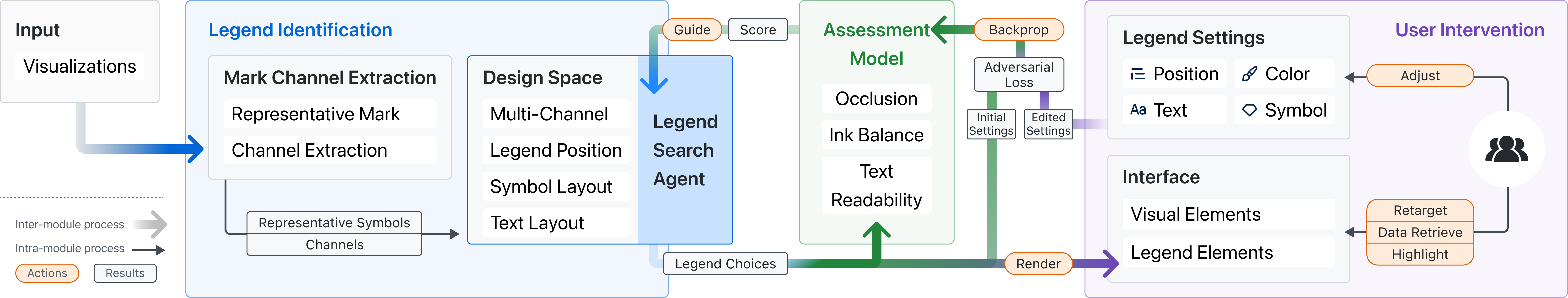}
     \caption{The workflow of AutoLegend begins with a visualization chart. After extracting representative markers and channels, the \textbf{legend search agent} conducts a solution search within the design space. The search process is guided by a quality \textbf{assessment model} that accepts evaluation metrics as input and outputs a scalar score for directing the legend search agent.
     The legend with the highest score is sent to the front-end interface, where users can interact and adjust its settings.
     The adjustments to the legend settings reflect the user's preferences, which are used to update the parameters of quality assessment model via real-time adversarial training.}
    \label{fig: pipeline}
  \end{figure*}

In recent years, many approaches have concentrated on automating various aspects of the visualization workflow, encompassing tasks such as recommending visualizations~\cite{hu2019vizml}, describing visualizations~\cite{liu2020autocaption}, answering questions about visualizations~\cite{kahou2017figureqa}, and generating automatic titles for visualizations~\cite{liu2023autotitle}.
These endeavors aim to enhance the accessibility of data and visualizations to a broader spectrum of users.
Despite the availability of many visualization tools that offer legend generation features, there remains a substantial number of visualizations that lack proper legends.
In light of this, we propose an automated approach for generating visualization legends tailored to a given visualization.
In \autoref{sec:design_space} and \autoref{sec: assess}, we summarize the design space of visualization legends and several factors that influence their quality.
However, while we recognize the importance of these factors, finding a good legend in a vast design space is not straightforward and can be challenging to explore manually.

\revision{Building upon the design space of legends summarized in \autoref{sec:design_space}, we propose a method for generating legends in real-time that adapts to user preferences. This automated approach is demonstrated using visualizations in SVG format.}
As shown in \autoref{fig: pipeline}, the automatic legend generation consists of four components: iconic symbol and channel extraction, the legend space search agent, a quality assessment model to guide the search agent, and an adversarial model that can adapt based on user input.
The \textbf{iconic symbol and channel extraction }component aims to identify the most representative visual elements and mappings from a given visualization.
The \textbf{legend search agent} searches for an optimal legend in the space of possible legends proposed in \autoref{sec:design_space}, with a focus on achieving high scores across multiple dimensions.
The \textbf{quality assessment model} guides the search agent by giving a score to legends searched by agent.
The \textbf{adversarial model} can adapt the search strategy based on the feedback provided by the user to improve the quality of the generated legends.

\subsection{Iconic Symbol and Channel Extraction}

\begin{figure*}[!ht]
    \centering
    \includegraphics[width=\textwidth]{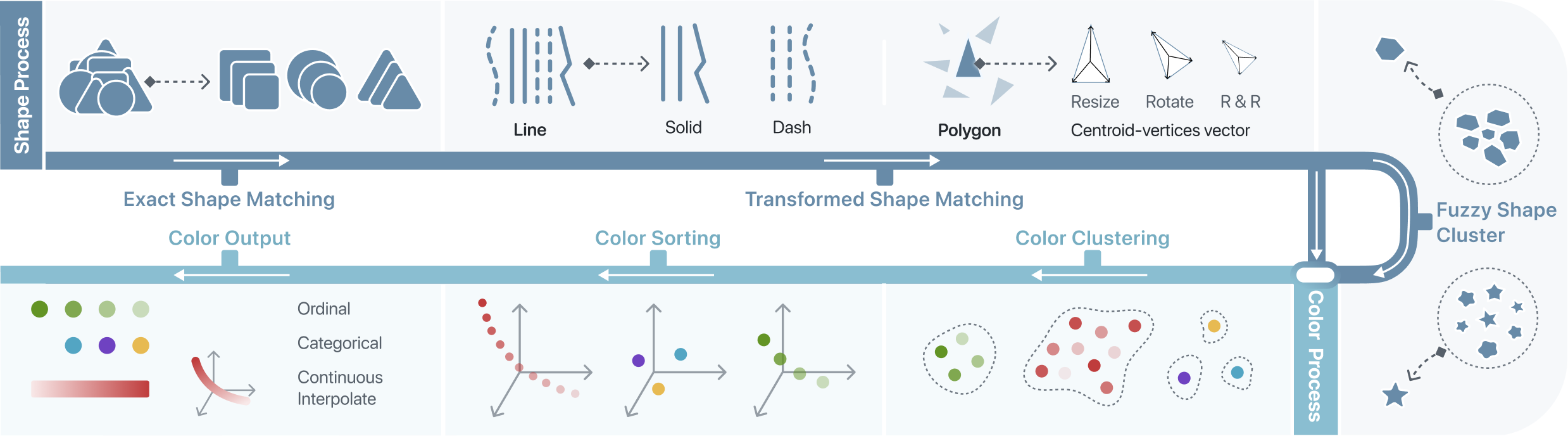}
    \caption{The process of legend parsing involves two main parts: shape extraction and color extraction. Simple to complex shape elements will be obtained by shape matching, rotation algorithms, and clustering analysis to derive representative graphics and their shape-related channels. Color extraction starts with color clustering, then sorting the colors and finally classifying them into continuous, sequential, or discrete categories.
    }
    \label{fig: extraction_channel}
\end{figure*}

This subsection aims to outline the extraction of iconic symbols in visualizations and inference of their information mapping channels. 
\revision{It is non-trivial to distinguish visual elements from messy DOM elements as well as extracting their data encoding channel. Take bar-chart as an example, some \textless{rect}\textgreater{} serves as bars while there might be other \textless{rect}\textgreater{} elements serve as background, axis. Also, some visualizations use \textless{path}\textgreater{} to render rectangular bars and circle instead of regular \textless{rect}\textgreater{} and \textless{circle}\textgreater{}. This makes the extraction more challenging and the pipeline should be able to distinguish them.}
The visual element extraction process comprises two primary steps: iconic symbol extraction and color recognition (\autoref{fig: extraction_channel}).
\revision{The first step involves the identification of the shape properties of visual elements, such as the element type, contour, and contour line width. A visualization contains lots of DOM elements, such as \textless{rect}\textgreater, \textless{path}\textgreater, \textless{circle}\textgreater, \textless{text}\textgreater. Some of them serves as axis, title, background while others serves as visual elements encoding information. This step decides which of those DOM elements encoding information, and divides these visual elements into groups based on their shape encoding channel. We call each group of visual elements as iconic symbol.}
In the second step, color recognition is employed to classify the colors within each group of iconic symbols. 
The color space may represent categorical, ordinal, or quantitative data attributes, and it is crucial to consider the information contained in the color mapping data during this step.

\textbf{Iconic Symbol Extraction.}
During this process, visualization symbols are identified and classified by their geometric features through a pipeline comprising three sequential steps. Three steps from simple to complex focus on the different geometric property. Once the visual elements are extracted, they serves as the input of color extraction process.

\begin{itemize}
    \item \textit{Exact Shape Matching.}
    In many visualizations, visual elements have exact same shapes, such as a bar chart using \textless{rect}\textgreater{} to represent all bars.
    We directly compare geometric contours to extract symbols with the same shapes.
    \item \textit{Transformed Shape Matching.}
    This step is dedicated to discovering transformed geometric shapes, including translation, rotation, and resizing.
    \revision{To remind that, DOM elements whose shape transformation can be easily extracted from ``transform'' attribute such as \textless{rect}\textgreater{} and \textless{text}\textgreater{} have been extracted in the previous step. 
    In this step, we mainly focus on \textless{path}\textgreater{}. Because the transformation might be directly mapped into the ``d'' attribute instead of shown explicitly in DOM's styles or attributes.}
    To achieve this, the centroid-vertices vector is used to characterize each shape, which is resistant to variations caused by rotation and scaling.
    The centroid-vertices vector is obtained by finding the centroid of a visualization glyph and then calculating the distances between the centroid and the vertices of the glyph.
    The vector is identical among shapes with different transformations.
    However, arcs and Bézier curves do not have vertices on the curve segment, so polygon approximation is applied to the curves to obtain the centroid-vertices vector.
    Finally, the centroid-vertices vector is normalized by dividing each entry by the maximum length.
    \item \textit{Fuzzy Shape Cluster.}
    \revision{In some scenarios, subtle disparities exist among \textless{path}\textgreater{}, such as every single region in voronoi diagram, different shape of mountains in \autoref{fig: cases} (e). We use shape clustering to find all these elements and select one representative shape from the cluster. If the user chose to generate semantic legends, our system will use this representative shape instead of normal squares as legends, presenting a semantic relationship with original visualization.}
    To accomplish this objective, we employ shape pattern clustering in a two-dimensional space utilizing area and aspect ratios of \textless{path}\textgreater{} elements.
    \revision{We use DBSCAN~\cite{ester1996dbscan} to cluster shapes and eliminate outliers and sparsely distributed shapes diverging from common shapes. Empirically, we set \(m\) to \(max(min(0.05 \times n, 20), 3)\), where \(n\) denotes the quantity of shapes, and \(\epsilon\) is set to 0.07. Each cluster denotes a shape pattern, and a random shape is designated as the representative for each cluster. Here, \(\epsilon\) defines the maximum distance between two points for one to be considered as in the neighborhood of the other, and \(m\) defines the minimum number of points required to form a dense region (cluster).}
\end{itemize}

\textbf{Color Mapping Extraction.}
The goal of color extraction is to find color encoding channel for each shape extracted previously. This process takes as input a collection of geometric shapes that were identified in the preceding shape extraction step. Our color extraction process also contains three steps. At any steps, shapes with different colors may be identified and outputted.

\begin{itemize}
\item \textit{Color Clustering.}
The geometric shape could carry multiple colors in a visualization.
This part aims to cluster colors of a shape into groups, including single-hue continuous color, multi-hue continuous, diverging continuous, ordinal, and categorical.
All RGB color values transform the CIELAB color space, followed by a normalization process.
These color points are subjected to clustering utilizing the DBSCAN~\cite{ester1996dbscan} algorithm within the normalized LAB color space.
The parameters governing this clustering process, denoted as \(m\) and \(\epsilon\), are assigned values of \(0.15\) and \(3\) respectively.
\item \textit{Color Ordering.}
For continuous legend, we obtain a cluster with several color samples. To build the continuous legend, these unordered color samples should be first sorted. 
This problem can be regarded as a traveling salesman problem (TSP). The goal is to find the shortest path that passes through all color points exactly once.
To this end, we employ an approximation algorithm. First, the minimum neighbor adjacency matrix is constructed using the nearest-neighbor algorithm. The distance calculations are made in CIELAB color space as well.
If the number of colors samples is vast (over 100), the minimum neighbor count will be increased to 3 to avoid disconnected adjacency graphs. As the starting and ending colors are unknown, we build a minimum spanning tree (MST) starting from each node and traverse it in preorder, obtaining and recording a sequence of color points and their corresponding costs. The sequence with the minimum cost is selected as the ordering of colors.
\item \textit{Color Interpolation.}
\revision{After sorting the color, the rearrange colors are interpolated to recover the original curve.} 
We utilize the cubic spline interpolation in the CIELAB space to fits a smooth curve. 
Then, we samples equidistant points on the curve. The number of sampled points is set to 512 by default. Since the interpolation is performed in the CIELAB space, the resulted continuous legends are more consistent with the human perception of colors.
\end{itemize}

\textbf{Post-Process.}
The same symbol can effectively convey information through multiple visual channels concurrently, with several channels potentially encoding identical data attributes.
For instance, the utilization of both size and color variations to represent a single data attribute serves as an illustrative example.
In the context of visual legend creation, it is customary to employ a single legend representation for content associated with the same data attribute.
To discern and extract these multi-channel mappings and correlations, a process of correlation analysis is undertaken across the various visual channels.
The correlations between the rotation angle, color, and scaling factor are calculated and analyzed. We use ``distance correlation''~\cite{distance_correlation} to measure the nonlinear correlation between multidimensional data.
This method is utilized to measure the correlations between the three-dimensional LAB color sequence, one-dimensional rotation angle sequence, and scaling factor sequence.
In practice, for any two of them, a nonlinear correlation is recognized if the distance vector exceeds a threshold of 0.75.

\subsection{Legend Search Agent}

After the iconic symbols and channel are extracted, there are two critical components involved: the legend search agent and the quality assessment model.
The responsibility of the search agent is to choose the most suitable legend solution in a high-dimensional mixed space.
This space includes numerous dimensions such as symbol layout (type and direction), text layout (type and color), multi-channel layout, and global layouts for legends.

The legend model agent navigate through this mixed high-dimensional space, which comprises both discrete and continuous dimensions. While the spatial position is a continuous dimension, others are discrete.
To search the legend space, we utilize a genetic algorithm~\cite{Mirjalili2019, gad2023pygad}, which is a typical mixed combinatorial optimization problem.
By simulating the process of visualization legend selection as genetic variation and receiving a score from the quality assessment model, the search agent generates better solutions over time. The quality assessment model plays a crucial role in guiding this process.

\subsection{Quality Assessment Model}

The model provides feedback on the exploration legend results generated by the search agent based on the multiple pre-defined indicators with default weights.
The user's feedback can update and adjust the weights in real time, which can be directly applied to the exploration model.
Our quality assessment model is a lightweight multi-layer perception that can be viewed as a two-layer, fully connected network.
The model takes the metrics mentioned in \autoref{sec: assess}
The weights of different indicators with different metrics mentioned are adjusted by simple annotations from users, which can embed expert knowledge and personalization settings. This quality assessment model, combined with the efficient exploration of the exploration model, can quickly respond to users' needs for creating, adjusting, and personalized recommending visualizations.

\begin{figure}[!ht]
    \centering
    \setlength{\belowcaptionskip}{-5px}
    \includegraphics[width=.7\columnwidth]{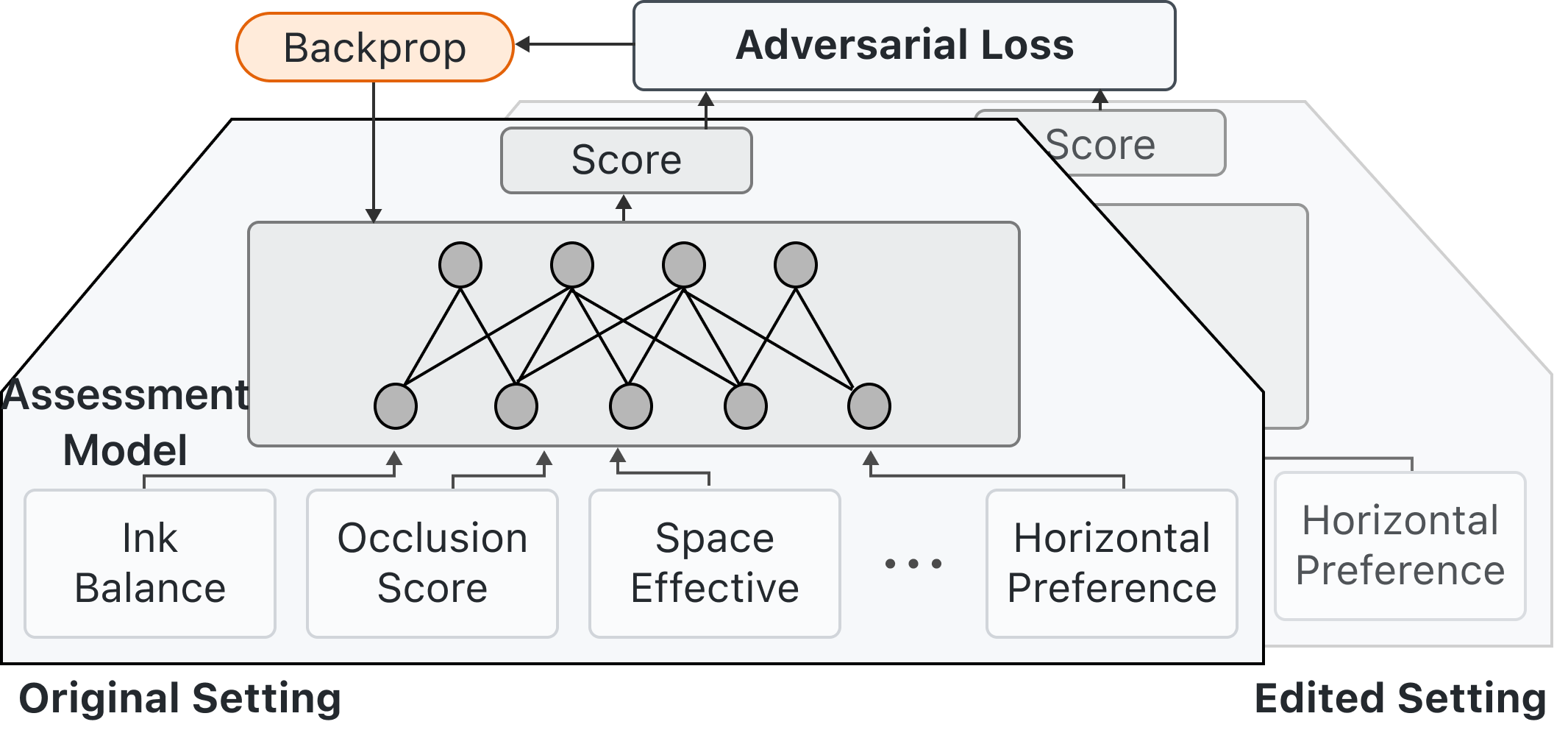}
    \caption{The quality assessment model and adversarial loss.
    The quality assessment model assesses the metrics of various legends and user preferences to compute a score. Conversely, the adversarial loss scrutinizes whether the partial order of the two components corresponds with the user's anticipations. Finally, the quality assessment model is back-propagated with the adversarial loss.
    }
    \label{fig: model}
\end{figure}

The quality assessment model takes several metrics as input, represented as $x$, namely, occlusion index, ink balance index, visualization bounding rectangle area index, horizontal preference, and vertical preference. 
It outputs a score S to represent the current legend solution's score:
$$S=r_\theta([x_1,x_2, ...,x_n]).$$

In our approach, we utilize feedback loop propagation to implement an online learning quality assessment model.
This model takes the metrics mentioned in \autoref{sec: assess} as input and outputs a scalar score to represent the quality of the current visualization. The quality assessment model is a multilayer perceptron. Users can directly modify the legend settings on the interactive interface, including changing the legend's symbol, text arrangement, and placement. These changes represent the user's knowledge or preferences. We collect these edit records, which essentially represent the user's preference in the form of partial order relation tuples. Based on these tuples, we perform back-propagation on the quality assessment model, which will be trained in real-time due to the small model scale. The subsequent continued use will employ the updated model, allowing the model to constantly adapt to the user's preferences.

\subsection{Online Updating based on User Feedback}

We advocate for utilizing feedback loop propagation in our approach.
We have implemented an online learning quality assessment model that takes multiple indicators as inputs and outputs a score to represent the current selection's quality. The quality assessment model is a lightweight perceptron (i.e., a fully-connected network).
After the user edits the location and selection of the graph on the interactive interface, the backend collects and stores the editing records. Based on the editing records, we assume that the user's edited result is better than the previous result. Using this partial order relationship binary tuple, we conduct real-time adversarial training of the model to update the model parameters quickly. Within seconds, subsequent graph calculations will use the updated quality assessment model. The effectiveness of this approach has been verified by experiments.

Given a pair of user inputs, $X_0$ and $X_1$, suppose the user considers $X_i$ to be superior. That is, the user adjusts from $X_1$ to $X_0$.
We impose the following adversarial loss function on the quality assessment model:
$$loss(r_{\theta}) = E_{(x_0, x_1,i)\sim D}[log(\sigma(r_{\theta}(x_i)-r_\theta(x_{1-i})))].$$
This loss function describes a pairwise prediction task. The model needs to predict the relative order $i$ of the given tuple $(x_0, x_1)$. Specifically, if $i = 0$, it means that $x_0$ is preferred over $x_1$; if $i = 1$, it means that $x_1$ is preferred over $x_0$. The calculation of the loss function is based on the prediction result $r_\theta(x)$ of the model, where $r_\theta(x)$ represents the prediction result of input $x$.
For the given tuple $(x_0, x_1)$, the probability $p_i$ of the relative order $i$ can be calculated using the model's prediction results as: 
$$p_i = \sigma(r_{\theta}(x_i)-r_\theta(x_{1-i})),$$
where $\sigma(x)$ denotes the sigmoid function, mapping any real number to the interval $(0, 1)$. For $r_\theta(x_i) - r_\theta(x_{1-i})$, a larger value of $r$ indicates that the model is more confident that $x_i$ is preferred over $x_{1-i}$; conversely, a smaller value indicates that the model is more confident that $x_i$ is less preferred compared to $x_{1-i}$.
When users modify the legend, it implies that they have a better alternative for the current legend. We can obtain at least one pair of user preference tuples based on the user's selection.

\section{Interaction on Legends}
\label{sec: legendinter}

\begin{figure}[!tb]
    \centering
    \includegraphics[width=\columnwidth]{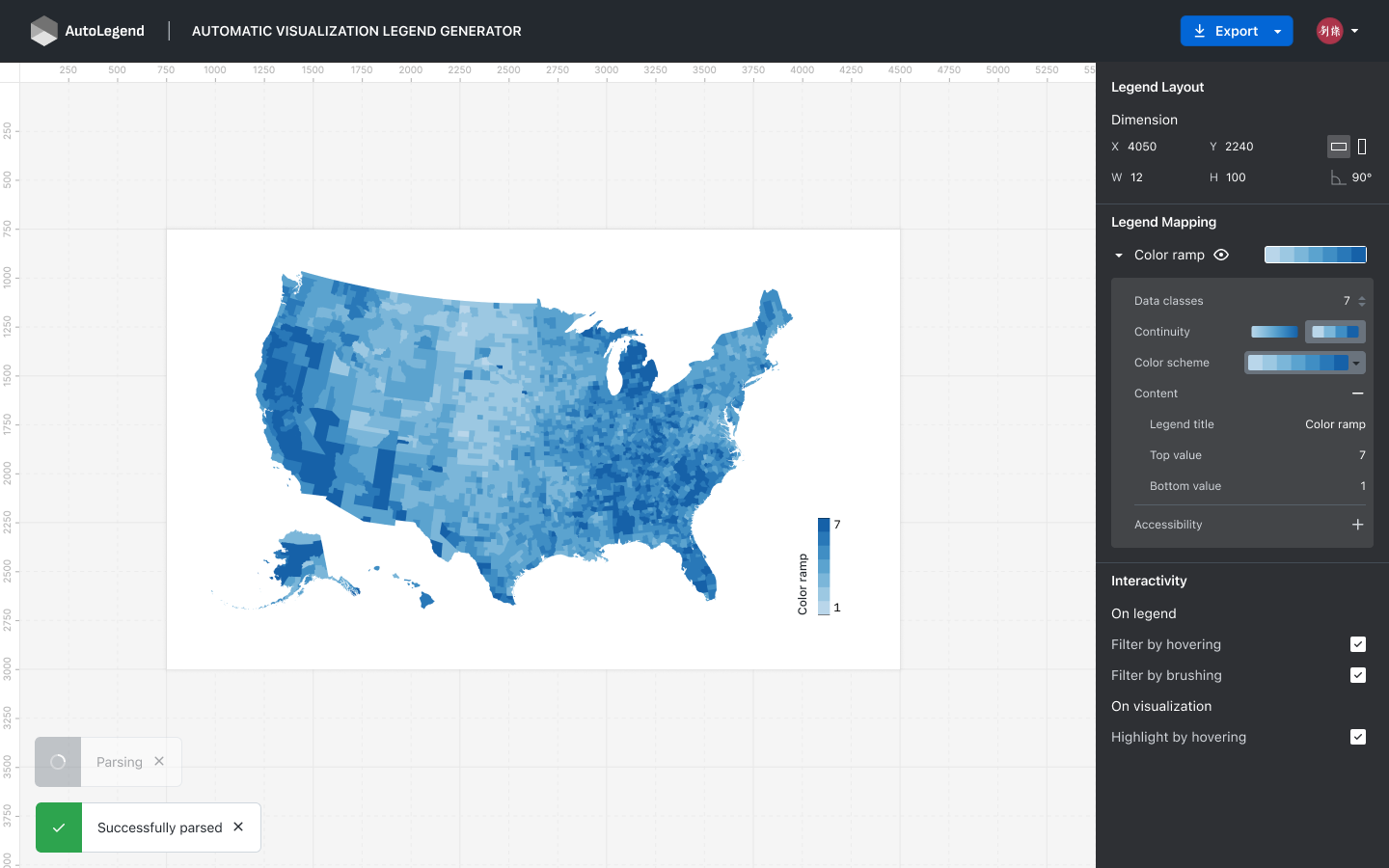}
    \caption{The user interface of AutoLegend includes the visualization view for placing the visualization and legend, the legend mapping view for users to provide mapping information, and the legend setting view for users to select and adjust legend settings.
   }
    \label{fig: interface}
  \end{figure}

Legends serve as a crucial bridge between data and visual marks in a visualization. After users upload a visualization to the interface (\autoref{fig: interface}), two natural types of interactions include legend-to-visualization interaction and visualization-to-legend interaction.
Legend-to-visualization interaction involves highlighting the corresponding parts in the visualization after the user selects them on the legend.
Visualization-to-legend interaction involves quickly retrieving data information on the legend after the user focuses on certain parts of the visualization.

\subsection{Legend-to-Visualization Interaction}

\begin{figure}[!tb]
    \centering
    \includegraphics[width=\columnwidth]{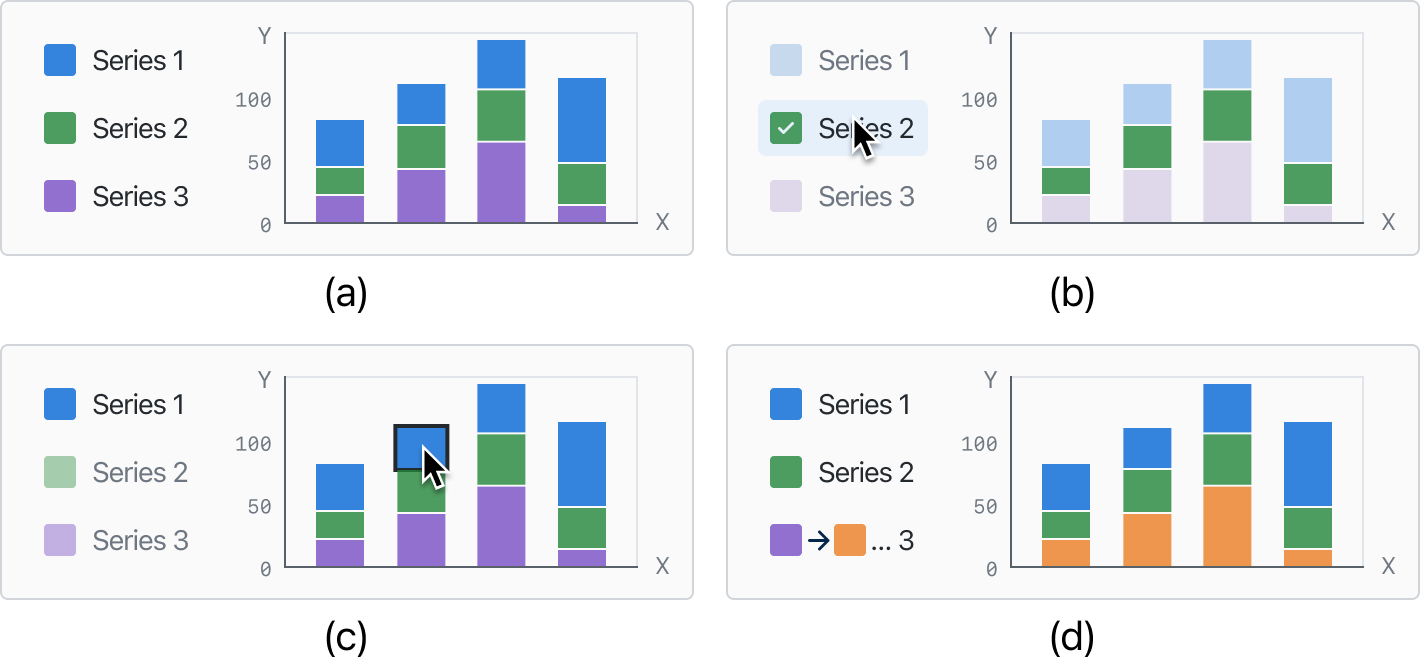}
    \caption{Interacting with a categorical legend. In Panel (a), we show the original visualization. In Panel (b), users can highlight a category in the legend, and the corresponding visualization elements will also be highlighted. In Panel (c), users can select visual elements, and the corresponding category in the legend will be highlighted. Finally, in Panel (d), users can modify the style of the visual elements, and the visualization content will be updated accordingly.
   }
    \label{fig: categorical_inter}
\end{figure}

\begin{figure}[!ht]
    \centering
    \includegraphics[width=\columnwidth]{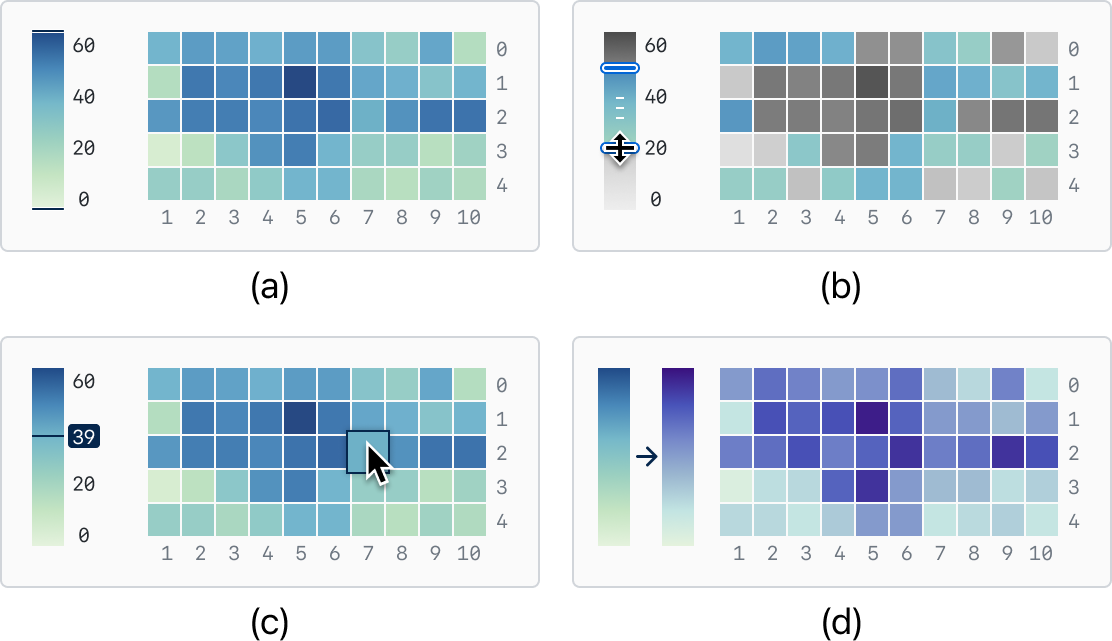}
    \caption{Interaction on continuous legends. (a) Original visualization. (b) Users can brush on the legend, and corresponding visual elements will be highlighted. (c) Users can select elements in the visualization, and the value in the legend will be highlighted. (d) Users can select a different style, and the visualization content will be retargeted accordingly.
    }
    \label{fig: continuous_inter}
\end{figure}

This section discusses the methods for legend-to-visualization interaction that enable users to select legend items and correspond them to visual elements in the visualization space. Discrete and continuous legends require different interaction methods. As shown in \autoref{fig: categorical_inter}, for a discrete legend, users typically select individual items of the legend. We provide this functionality through point-and-click selection on the legend to support user selection.
For continuous legends, as shown in \autoref{fig: continuous_inter}, we represent the selection of continuous ranges in several different states. We support specifying upper and lower bounds with specific numerical values or ranges composed of them.

Furthermore, the legend can also serve as a means for retargeting visualizations.
The mapping relationship extracted from the visualization contains information about the original visualization elements and their corresponding attribute values.
Our legend supports the modification of the original visualization elements, including but not limited to changing color channels and stroke width. \autoref{fig: continuous_inter} (d) illustrates our transformation from a blue color scheme to a purple color scheme.

\subsection{Visualization-to-Legend Interaction}

Visualization-to-legend interaction supports obtaining response and data information on the legend when the user focuses on certain parts of the visualization that represent data. After the user hovers or clicks on visual marks in the visualization, we enhance and highlight the corresponding data on the legend through statistical data enhancement and highlighting.

\begin{figure*}[!ht]
    \centering
    \includegraphics[width=\textwidth]{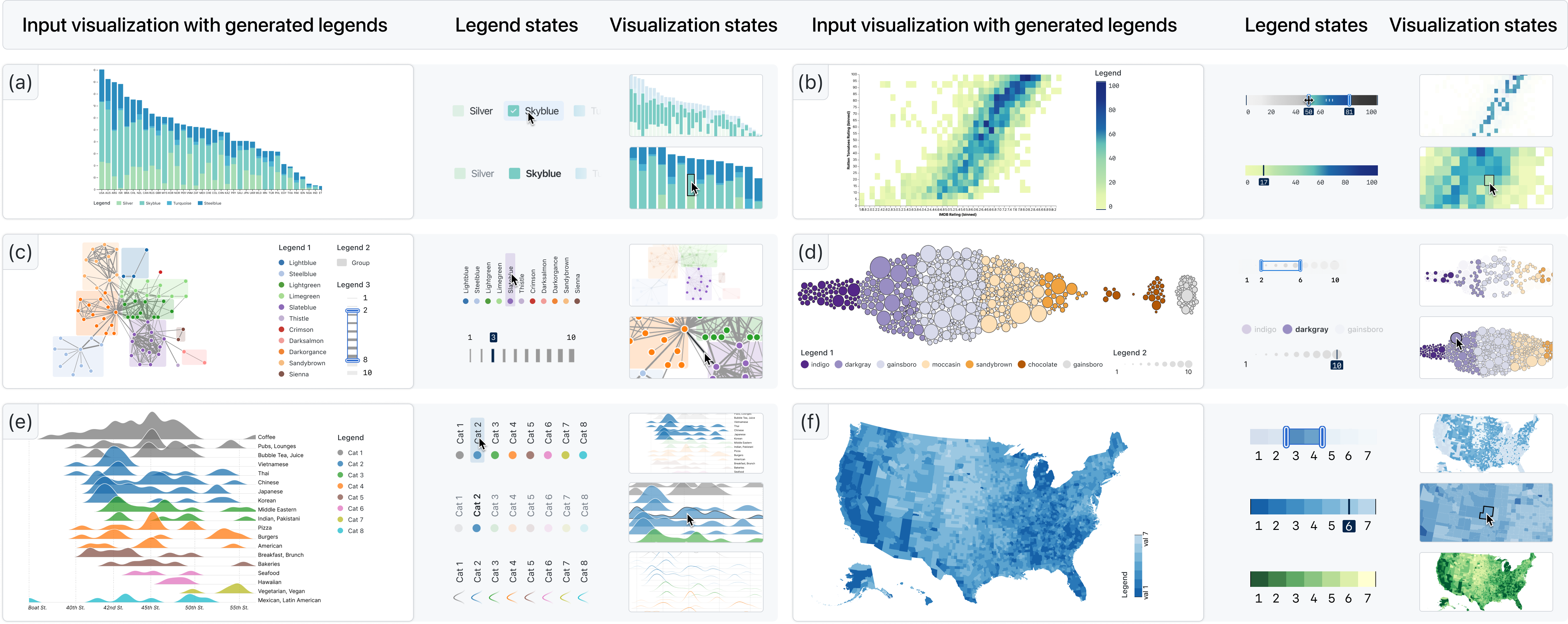}
    \caption{AutoLegend generates legends for a variety of visualizations.
    This tool takes into account the appropriate position and settings for the legend generation. The legends support users interacting with the visualization.}
    \label{fig: cases}
\end{figure*}

\section{Cases}
\label{sec: cases}

In this section, we demonstrate six visualization legends generated by AutoLegend, including bar charts, network diagrams, maps, area charts, heat maps, and bubble charts. These visualizations encompass various channels, such as categorical, ordinal, and continuous. They also include legends for single visualization channels as well as multiple visualization channels.

\subsection{Single-channel Legends}

\begin{figure}[!ht]
    \centering
    \includegraphics[width=\columnwidth]{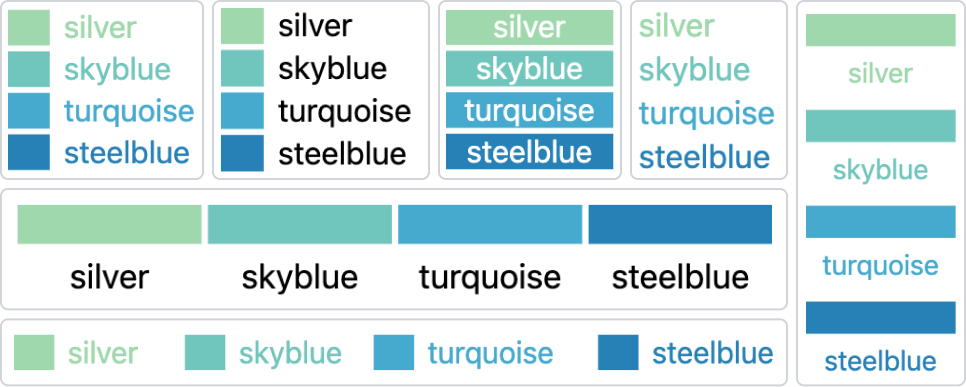}
    \caption{Legends generated for the stacked bar chart of \autoref{fig: cases} (e).
    }
    \label{fig: generatedLegend}
\end{figure}

AutoLegend can easily handle single-channel visualizations, as demonstrated in \autoref{fig: cases} (a), (b), (e), and (f), which feature discrete, ordinal, and continuous color channels, respectively.
These types of charts comprise a significant portion of visualizations, and AutoLegend can generate corresponding legends and choose appropriate positions for them.
AutoLegend can generate various legends for a visualization, as shown in \autoref{fig: generatedLegend}, which are generated legends for the stacked bar chart in \autoref{fig: cases} (a).

AutoLegend also enables users to interact with the legends in three ways: retrieving corresponding values on the legend through the visualization, highlighting corresponding visual objects in the visualization through the legend, or retargeting the visualization through the legend to change its style.
For instance, users can interact with the legend to select the focus categories, which will highlight the corresponding bars in the chart. As demonstrated in \autoref{fig: cases} (e), users can retarget an area chart to a line chart by setting the stroke color according to its fill color and setting the fill to none. \autoref{fig: cases} (f) illustrates how the visualization can be retargeted to a new color scheme.

\subsection{Multi-channel Legends}
\label{subsec:complex_discrete_color_visualizations}

\autoref{fig: cases} (f) depicts a visualization with elements that have both size and color attributes. After extracting the two channels, AutoLegend generates two separate legends representing the categorical colors and sizes of the circles. These legends also support user interactions.
\autoref{fig: cases} (e) displays a visualization with elements featuring multiple different marks; AutoLegend can generate distinct legends for circles, rectangles, and lines. Corresponding interactions are also facilitated on these legends. The legend maps node colors to different categories and the thickness of the link to the connections of the nodes.
Users can interact with the legend to highlight nodes or edges based on specific attributes.

\section{User Study}
\label{sec: study}

We aim to explore whether AutoLegend can effectively generate legends and improve based on user usage.
\revision{This user study aims to achieve two objectives: firstly, to assess the effectiveness of generative legends, and secondly, to validate the efficacy of the adaptive learning algorithm in aligning with user preferences.
This is particularly pertinent since the  undergoes continuous updates during user interactions.
}

\subsection{Study Design}

\par
\textbf{Participants:} We recruited 13 participants, four of whom were female.
The participants included 2 undergraduate students and 11 graduate students. We collected information on their academic backgrounds and experience with visualization using a 5-point Likert scale. The results showed that most of the participants had prior experience using programming tools ($\mu=4.62$, $\sigma=0.87$).
All participants agreed to have their usage records collected.

\textbf{Procedure:} We introduced the system to the participants and demonstrated how to upload visualizations, generate legends, modify legends, and interact with them. We selected 24 representative visualizations from the galleries of EChart, D3, and Vega-Lite, including common types such as bar charts, choropleth maps, heatmaps, and node-link graphs. The cases discussed in \autoref{sec: cases} were also included. Subsequently, we used 18 of these visualizations for the participants to interact with the system and generate legends.
During the legend generation process, users are allowed to edit the settings of visualization legends, such as altering the position, symbol layout, and text layout, if they deemed there was a better choice.
The system continuously collected user feedback and updated the model in real-time.
AutoLegend generated an adapted for each user, and both the adapted and original produced visualization legends for the remaining 6 visualizations.
To validate AutoLegend's capability to learn user preferences, legends generated by two models were presented to users for evaluation, without revealing which legends were produced by the customized model.

\textbf{Interview and Questionnaire:} After the legend generation process, we conducted interviews with the participants to gauge their perceptions of the system's interactivity and the quality of the generated legends. The interviews encouraged participants to freely discuss their thoughts about the system. Following the interview, participants were asked to rate the accuracy of the extracted mark and channel, the usability of the legend interactions, and the editing function of the system.
Specifically, participants were asked to evaluate the usability of legend interactions, such as highlighting and filtering, as well as the effectiveness of the legends in redirecting visual content. These aspects were scored on a scale of 1 to 5, where 1 indicated ``very useless'' and 5 indicated ``very effective''.
Lastly, we asked participants to rate the generated legends from both the original and adapted legend on a scale of 1 to 5 points.

\subsection{Feedback}
In summary, AutoLegend has been described by participants as having a rich variety of legend generation, being highly interactive, and being very helpful.
We summarize the user feedback in the following texts.

\begin{itemize}[leftmargin=*]
    \item \textit{Effective Preference Acquisition.}
    In addition to the aforementioned 18 visualizations for user interaction and training purposes, we conducted an evaluation on the remaining 6 visualizations. These visualizations encompass stacked bar charts, line plots, maps, pie charts, and bar charts. For each individual user, our system is capable of producing an adapted model following user preference. Employing both the Adapted model and the system's default parameter settings, we generated paired sets of legends for the aforementioned 6 visualizations. Subsequently, each user rated their level of preference for the 12 visualization legends on a scale of 1 to 5, with 1 indicating the least preferred and 5 indicating the most preferred.
    13 participants in total contributed 156 ratings across both the default and preference-adjusted styles. To ascertain whether a significant difference in preference exists between the legends generated by default and those adjusted according to user preferences, a two-sample t-test was employed. The experimental results revealed that the preference-adjusted legends (\(\mu = 4.13\), \(\sigma = 0.93\)) received significantly higher ratings compared to the default settings (\(\mu = 3.79\), \(\sigma = 0.97\)). The calculated two-sided t-value was 2.38, resulting in a p-value of 0.0195, which is less than the significance threshold of 0.05. As such, a statistically significant difference between the two sets of legends is confirmed.
    Although we did not explicitly inform users about the acquisition of their preferences, our results demonstrate that users exhibited certain preferences in legend selection, and AutoLegend successfully captured these preferences. For example, P4 tended to choose legends located closer to the right edges, while P6 favored legends positioned closer to the middle bottom.
    \item \textit{Accurate Encoding Channel Identification.}
    AutoLegend is capable of extracting various channels from visualization accurately and determining whether these channels map to data ($\mu=4.54$, $\sigma=0.52$).
    This allows for the generation of diverse legends with semantic connections to the original visual elements.
    For example, P2 expressed appreciation for the preserved contour in the ``mountain'' visualization legend. Moreover, AutoLegend can detect correlations between different channels. One user was impressed by the system's ability to uncover the relationship between the color and angle of the ``wind'' element and reflect it in the legend, a mapping relationship they found challenging to identify themselves.
    Another user praised the system's comprehensiveness and correspondence in extracting multiple types of elements, such as extracting lines of varying colors and widths, as well as differently sized polygons.
    \item \textit{Diverse Interactions and Effective Redirection.}
    Participants regard AutoLegend can generate various legends ($\mu=4.31$, $\sigma=0.75$).
    AutoLegend offers bidirectional interactions between visualizations and legends, including
    highlighting corresponding legends when hovering over visualization elements, highlighting corresponding visualization elements when hovering over legends, continuous legend filtering and highlighting, and providing data reminders for continuous channel elements. Many users noted the diverse range of interactions offered by AutoLegend, which aided their understanding of the visualizations. Additionally, users expressed a strong interest in redirected visualizations. All users were eager to experiment with changing visualization colors through legends and found the ability to ``modify data mappings in legends" very useful.
    
\end{itemize}

\section{Discussions and Future Works}
\label{sec: discuss}
In this section, we delve into the insights gained from our research, as well as identify potential avenues for future work.
\remark{But demonstrating the a few failure case will help the community to show the limitation and the future work. }

\subsection{Editing Sub-Legends}
In the interviews, users mentioned the desirability of allowing the manipulation of sub-legends within visualizations, such as merging or deleting specific elements.
For instance, in some visualizations, users might not want to display certain channels. These editing operations should be integrated into the editing process for visualization legends. In the future, we aim to enable the model to learn user preferences for editing various channels, rather than merely providing the editing functionality. This will require incorporating the relationships between legends into the representation of user preferences.

\subsection{Binding External Data}
Within our current framework, we accept a visualization as input and extract symbols and channels that map data within the visualization. While this approach has broad applications, some users have expressed interest in supporting the binding of external data, which would allow for the addition of more information to the original visualization. For example, in a node-link diagram, the original visualization might not have size mapped to any other data. However, now we can allow the size to represent additional attributes, thus enhancing the retargeting capabilities of the system.

\subsection{Intelligent Interaction on Legend}

Currently, our model is capable of providing interactive visualization through the use of legends.
This includes highlighting corresponding visual elements. Some participants have suggested that there may be deeper data relationships between sub-legends that could be identified and highlighted simultaneously.
By doing so, we could provide a more informative experience for users.
For example, in a node-link diagram, highlighting a node and its corresponding edges at the same time would provide a more comprehensive understanding of the underlying data relationships.

In the future, we plan to construct intelligent connections by parsing and identifying common patterns in the data.
This will allow us to highlight relevant visualization components in a more meaningful way.
By incorporating this level of intelligence into the interaction process, we can improve the user experience and facilitate a more comprehensive understanding of the underlying data relationships within the visualizations.

\subsection{Integrating Human-AI Collaboration}

In traditional machine learning approaches, data annotation and model training are often treated as separate processes. It is common for the individuals responsible for data annotation, quality control, and model training to differ from those who utilize the model in interactive scenarios. As a result, user interactions typically rely on a fixed, pre-trained model, making it challenging to integrate user feedback into the model effectively. A significant loop is required for user feedback during the interaction process to provide input to the model.

This process entails the user establishing reasonable feedback rules, ensuring data compliance, and training the model to incrementally incorporate user feedback while preserving previously learned information. Additionally, the model trainer must undergo training, and the model must be redeployed in the interactive environment. However, the duration of this loop may exceed the allowable interaction delay, causing users to lack a continuous interaction process. This procedure may necessitate collaboration among multiple parties.

In our comprehensive loop, interaction, learning, and annotation should form a cohesive cycle, with interaction acting as an annotation. This process offers valuable learning examples, enabling the machine to continually learn from user feedback.

\section{Conclusion}
\label{sec: conclusion}

In this paper, we present a novel approach for creating effective visualization legends that can adapt to user preferences. To achieve this goal, we first summarize the design space of the visualization legend, such as visual channels, visual marks, element layout, text layout, and multi-channel layout. We then created evaluation metrics for each of these design elements to help determine their effectiveness.
Our method generates a legend that meets the established metrics by searching through a high-dimensional space of potential legend designs and scoring them based on the metrics. We also incorporated a preference metric that considers user feedback, allowing for real-time adjustments to the user interaction and dynamic updates to the back-end model.

\bibliographystyle{IEEEtran}
\bibliography{manuscript}

\begin{thebibliography}{10}
\providecommand{\url}[1]{#1}
\csname url@samestyle\endcsname
\providecommand{\newblock}{\relax}
\providecommand{\bibinfo}[2]{#2}
\providecommand{\BIBentrySTDinterwordspacing}{\spaceskip=0pt\relax}
\providecommand{\BIBentryALTinterwordstretchfactor}{4}
\providecommand{\BIBentryALTinterwordspacing}{\spaceskip=\fontdimen2\font plus
\BIBentryALTinterwordstretchfactor\fontdimen3\font minus
  \fontdimen4\font\relax}
\providecommand{\BIBforeignlanguage}[2]{{%
\expandafter\ifx\csname l@#1\endcsname\relax
\typeout{** WARNING: IEEEtran.bst: No hyphenation pattern has been}%
\typeout{** loaded for the language `#1'. Using the pattern for}%
\typeout{** the default language instead.}%
\else
\language=\csname l@#1\endcsname
\fi
#2}}
\providecommand{\BIBdecl}{\relax}
\BIBdecl

\bibitem{bostock2011d3}
M.~Bostock, V.~Ogievetsky, and J.~Heer, ``D$^3$: {Data-Driven Documents},''
  \emph{IEEE Trans. Vis. Comput. Graph.}, vol.~17, no.~12, pp. 2301--2309, Dec.
  2011.

\bibitem{satyanarayan2017vegalite}
A.~{Satyanarayan}, D.~{Moritz}, K.~{Wongsuphasawat}, and J.~{Heer},
  ``Vega-lite: A grammar of interactive graphics,'' \emph{IEEE Trans. Vis.
  Comp. Graph.}, vol.~23, no.~1, pp. 341--350, 2017.

\bibitem{li2018echarts}
D.~Li, H.~Mei, Y.~Shen, S.~Su, W.~Zhang, J.~Wang, M.~Zu, and W.~Chen,
  ``{ECharts}: A declarative framework for rapid construction of web-based
  visualization,'' \emph{Visual Informatics}, vol.~2, no.~2, pp. 136--146,
  2018.

\bibitem{sieber2005smart}
R.~Sieber, C.~Schmid, and S.~Wiesmann, ``Smart legend--{Smart} atlas,'' in
  \emph{Proceedings of International Conference of the ICA}, 2005.

\bibitem{gobel2018gazelegend}
F.~G\"{o}bel, P.~Kiefer, I.~Giannopoulos, A.~T. Duchowski, and M.~Raubal,
  ``Improving map reading with gaze-adaptive legends,'' in \emph{Proceedings of
  the ACM Symposium on Eye Tracking Research and Applications}, 2018.

\bibitem{dykes2010rethinkmaplegend}
J.~Dykes, J.~Wood, and A.~Slingsby, ``Rethinking map legends with
  visualization,'' \emph{IEEE Trans. Vis. Comput. Graph.}, vol.~16, no.~6, pp.
  890--899, 2010.

\bibitem{imhof1972thematische}
E.~Imhof, \emph{Thematische Kartographie}.\hskip 1em plus 0.5em minus
  0.4em\relax Berlin: Walter de Gruyter, 1972.

\bibitem{edler2020searching}
D.~Edler, J.~Keil, M.-C. Tuller, A.-K. Bestgen, and F.~Dickmann, ``Searching
  for the ‘right’legend: the impact of legend position on legend decoding
  in a cartographic memory task,'' \emph{The Cartographic Journal}, vol.~57,
  no.~1, pp. 6--17, 2020.

\bibitem{riche2010understanding}
N.~H. Riche, B.~Lee, and C.~Plaisant, ``Understanding interactive legends: a
  comparative evaluation with standard widgets,'' \emph{Computer Graphics
  Forum}, vol.~29, no.~3, pp. 1193--1202, 2010.

\bibitem{shahira2021towards}
K.~Shahira and A.~Lijiya, ``Towards assisting the visually impaired: a review
  on techniques for decoding the visual data from chart images,'' \emph{IEEE
  Access}, vol.~9, pp. 52\,926--52\,943, 2021.

\bibitem{Vis4nonvisual}
J.~Choi, S.~Jung, D.~G. Park, J.~Choo, and N.~Elmqvist, ``Visualizing for the
  non-visual: Enabling the visually impaired to use visualization,''
  \emph{Computer Graphics Forum}, vol.~38, no.~3, pp. 249--260, 2019.

\bibitem{savva2011revision}
M.~Savva, N.~Kong, A.~Chhajta, F.-F. Li, M.~Agrawala, and J.~Heer, ``{ReVision:
  Automated classification, analysis and redesign of chart images},'' in
  \emph{Proceedings of ACM UIST}, 2011, pp. 393--402.

\bibitem{poco2017reverse}
J.~Poco and J.~Heer, ``Reverse-engineering visualizations: Recovering visual
  encodings from chart images,'' \emph{Computer Graphics Forum}, vol.~36,
  no.~3, pp. 353--363, 2017.

\bibitem{yuan2022colormap}
L.-P. Yuan, W.~Zeng, S.~Fu, Z.~Zeng, H.~Li, C.-W. Fu, and H.~Qu, ``Deep
  colormap extraction from visualizations,'' \emph{IEEE Trans. Vis. Comput.
  Graph.}, vol.~28, no.~12, pp. 4048--4060, 2022.

\bibitem{Luo2021WACV}
J.~Luo, Z.~Li, J.~Wang, and C.-Y. Lin, ``{ChartOCR}: Data extraction from
  charts images via a deep hybrid framework,'' in \emph{Proceedings of IEEE
  WACV}, 2021, pp. 1917--1925.

\bibitem{Lai2020Annotation}
C.~Lai, Z.~Lin, R.~Jiang, Y.~Han, C.~Liu, and X.~Yuan, ``Automatic annotation
  synchronizing with textual description for visualization,'' in
  \emph{Proceedings of the SIGCHI Conference on Human Factors in Computing
  Systems}, paper No. 316, 2020.

\bibitem{zhou2021reverse}
F.~Zhou, Y.~Zhao, W.~Chen, Y.~Tan, Y.~Xu, Y.~Chen, C.~Liu, and Y.~Zhao,
  ``Reverse-engineering bar charts using neural networks,'' \emph{Journal of
  Visualization}, vol.~24, pp. 419--435, 2021.

\bibitem{Zhang2020MI3}
Y.~Zhang, M.~Chen, and B.~Coecke, ``{MI3: Machine-Initiated Intelligent
  Interaction for Interactive Classification and Data Reconstruction},''
  \emph{ACM Trans. Interact. Intell. Syst.}, vol.~11, no. 3–4, 2021.

\bibitem{liu2019data}
X.~Liu, D.~Klabjan, and P.~NBless, ``Data extraction from charts via single
  deep neural network,'' \emph{arXiv preprint arXiv:1906.11906}, 2019.

\bibitem{Poco2018extract}
J.~Poco, A.~Mayhua, and J.~Heer, ``Extracting and retargeting color mappings
  from bitmap images of visualizations,'' \emph{IEEE Trans. Vis. Comput.
  Graph.}, vol.~24, no.~1, pp. 637--646, 2018.

\bibitem{hoque2019searching}
E.~Hoque and M.~Agrawala, ``Searching the visual style and structure of d3
  visualizations,'' \emph{IEEE Trans. Vis. Comput. Graph.}, vol.~26, no.~1, pp.
  1236--1245, 2019.

\bibitem{cui2021mixed}
W.~Cui, J.~Wang, H.~Huang, Y.~Wang, C.-Y. Lin, H.~Zhang, and D.~Zhang, ``A
  mixed-initiative approach to reusing infographic charts,'' \emph{IEEE Trans.
  Vis. Comput. Graph.}, vol.~28, no.~1, pp. 173--183, 2021.

\bibitem{Chen2020timeline}
Z.~Chen, Y.~Wang, Q.~Wang, Y.~Wang, and H.~Qu, ``Towards automated infographic
  design: Deep learning-based auto-extraction of extensible timeline,''
  \emph{IEEE Trans. Vis. Comput. Graph.}, vol.~26, no.~1, pp. 917--926, 2019.

\bibitem{kong2012graphical}
N.~Kong and M.~Agrawala, ``Graphical overlays: Using layered elements to aid
  chart reading,'' \emph{IEEE Trans. Vis. Comput. Graph.}, vol.~18, no.~12, pp.
  2631--2638, 2012.

\bibitem{Lu2021ant}
M.~Lu, N.~Fish, S.~Wang, J.~Lanir, D.~Cohen-Or, and H.~Huang, ``Enhancing
  static charts with data-driven animations,'' \emph{IEEE Trans. Vis. Comput.
  Graph.}, vol.~28, no.~7, pp. 2628--2640, 2022.

\bibitem{choi2015visdock}
J.~Choi, D.~G. Park, Y.~L. Wong, E.~Fisher, and N.~Elmqvist, ``{VisDock}: A
  toolkit for cross-cutting interactions in visualization,'' \emph{IEEE Trans.
  Vis. Comput. Graph.}, vol.~21, no.~9, pp. 1087--1100, 2015.

\bibitem{harper2014deconstructing}
J.~Harper and M.~Agrawala, ``Deconstructing and restyling {D$^3$}
  visualizations,'' in \emph{Proceedings of ACM UIST}, 2014, pp. 253--262.

\bibitem{harper2018converting}
------, ``Converting basic {D3} charts into reusable style templates,''
  \emph{IEEE Trans. Vis. Comput. Graph.}, vol.~24, no.~3, pp. 1274--1286, 2018.

\bibitem{lu2017interaction}
M.~Lu, J.~Liang, Y.~Zhang, G.~Li, S.~Chen, Z.~Li, and X.~Yuan, ``Interaction+:
  Interaction enhancement for web-based visualizations,'' in \emph{Proceedings
  of the IEEE Pacific Visualization Symposium}, 2017, pp. 61--70.

\bibitem{liu2024spatial}
C.~Liu, Y.~Zhang, C.~Wu, C.~Li, and X.~Yuan, ``A spatial constraint model for
  manipulating static visualizations,'' \emph{ACM Transactions on Interactive
  Intelligent Systems}, vol.~14, no.~2, pp. 1--29, 2024.

\bibitem{goodcase_a}
B.~Alper, N.~Riche, G.~Ramos, and M.~Czerwinski, ``Design study of linesets, a
  novel set visualization technique,'' \emph{IEEE Trans. Vis. Comput. Graph.},
  vol.~17, no.~12, pp. 2259--2267, 2011.

\bibitem{goodcase_b}
F.~Chevalier, R.~Vuillemot, and G.~Gali, ``Using concrete scales: A practical
  framework for effective visual depiction of complex measures,'' \emph{IEEE
  Trans. Vis. Comput. Graph.}, vol.~19, no.~12, pp. 2426--2435, 2013.

\bibitem{goodcase_c}
Y.~Yang, T.~Dwyer, S.~Goodwin, and K.~Marriott, ``Many-to-many
  geographically-embedded flow visualisation: An evaluation,'' \emph{IEEE
  Trans. Vis. Comput. Graph.}, vol.~23, no.~1, pp. 411--420, 2017.

\bibitem{goodcase_d_up}
C.~Palomo, Z.~Guo, C.~T. Silva, and J.~Freire, ``Visually exploring
  transportation schedules,'' \emph{IEEE Trans. Vis. Comput. Graph.}, vol.~22,
  no.~1, pp. 170--179, 2016.

\bibitem{goodcase_d_down}
M.~Rubio-Sánchez and A.~Sanchez, ``Axis calibration for improving data
  attribute estimation in star coordinates plots,'' \emph{IEEE Trans. Vis.
  Comput. Graph.}, vol.~20, no.~12, pp. 2013--2022, 2014.

\bibitem{goodcase_e_up}
M.~Jarema, I.~Demir, J.~Kehrer, and R.~Westermann, ``Comparative visual
  analysis of vector field ensembles,'' in \emph{Proceedings of IEEE VAST},
  2015, pp. 81--88.

\bibitem{goodcase_e_middle_down}
Y.~Lu, M.~Steptoe, S.~Burke, H.~Wang, J.-Y. Tsai, H.~Davulcu, D.~Montgomery,
  S.~R. Corman, and R.~Maciejewski, ``Exploring evolving media discourse
  through event cueing,'' \emph{IEEE Trans. Vis. Comput. Graph.}, vol.~22,
  no.~1, pp. 220--229, 2016.

\bibitem{goodcase_f}
W.~Dou, C.~Ziemkiewicz, L.~Harrison, D.~H. Jeong, R.~Ryan, W.~Ribarsky,
  X.~Wang, and R.~Chang, ``Comparing different levels of interaction
  constraints for deriving visual problem isomorphs,'' in \emph{Proceedings of
  IEEE VAST}, 2010, pp. 195--202.

\bibitem{goodcase_g}
J.~Zhao, F.~Chevalier, C.~Collins, and R.~Balakrishnan, ``Facilitating
  discourse analysis with interactive visualization,'' \emph{IEEE Trans. Vis.
  Comput. Graph.}, vol.~18, no.~12, pp. 2639--2648, 2012.

\bibitem{goodcase_h}
W.~Meulemans, J.~Dykes, A.~Slingsby, C.~Turkay, and J.~Wood, ``Small multiples
  with gaps,'' \emph{IEEE Trans. Vis. Comput. Graph.}, vol.~23, no.~1, pp.
  381--390, 2017.

\bibitem{goodcase_i}
R.~Scheepens, N.~Willems, H.~van~de Wetering, G.~Andrienko, N.~Andrienko, and
  J.~J. van Wijk, ``Composite density maps for multivariate trajectories,''
  \emph{IEEE Trans. Vis. Comput. Graph.}, vol.~17, no.~12, pp. 2518--2527,
  2011.

\bibitem{goodcase_j}
P.~Bak, F.~Mansmann, H.~Janetzko, and D.~Keim, ``Spatiotemporal analysis of
  sensor logs using growth ring maps,'' \emph{IEEE Trans. Vis. Comput. Graph.},
  vol.~15, no.~6, pp. 913--920, 2009.

\bibitem{goodcase_k}
J.~Wood, A.~Slingsby, N.~Khalili-Shavarini, J.~Dykes, and D.~Mountain,
  ``Visualization of uncertainty and analysis of geographical data,'' in
  \emph{Proceedings of IEEE VAST}, 2009, pp. 261--262.

\bibitem{goodcase_l}
X.~Yuan, P.~Guo, H.~Xiao, H.~Zhou, and H.~Qu, ``Scattering points in parallel
  coordinates,'' \emph{IEEE Trans. Vis. Comput. Graph.}, vol.~15, no.~6, pp.
  1001--1008, 2009.

\bibitem{deng2022visimages}
D.~Deng, Y.~Wu, X.~Shu, J.~Wu, S.~Fu, W.~Cui, and Y.~Wu, ``{VisImages}: A
  fine-grained expert-annotated visualization dataset,'' \emph{IEEE Trans. Vis.
  Comput. Graph.}, no.~01, pp. 1--1, 2022.

\bibitem{hu2019vizml}
K.~Hu, M.~A. Bakker, S.~Li, T.~Kraska, and C.~Hidalgo, ``{VizML}: A machine
  learning approach to visualization recommendation,'' in \emph{Proceedings of
  the SIGCHI Conference on Human Factors in Computing Systems}, 2019.

\bibitem{liu2020autocaption}
C.~Liu, L.~Xie, Y.~Han, X.~Yuan \emph{et~al.}, ``Autocaption: An approach to
  generate natural language description from visualization automatically,'' in
  \emph{Proceedings of the IEEE Pacific Visualization Symposium (Notes)}, 2020,
  pp. 191--195.

\bibitem{kahou2017figureqa}
S.~E. Kahou, V.~Michalski, A.~Atkinson, {\'A}.~K{\'a}d{\'a}r, A.~Trischler, and
  Y.~Bengio, ``Figureqa: An annotated figure dataset for visual reasoning,''
  \emph{arXiv preprint arXiv:1710.07300}, 2017.

\bibitem{liu2023autotitle}
C.~Liu, Y.~Guo, and X.~Yuan, ``{AutoTitle}: An interactive title generator for
  visualizations,'' \emph{IEEE Trans. Vis. Comp. Graph.}, pp. 1--12, 2023.

\bibitem{ester1996dbscan}
M.~Ester, H.-P. Kriegel, J.~Sander, X.~Xu \emph{et~al.}, ``A density-based
  algorithm for discovering clusters in large spatial databases with noise,''
  in \emph{KDD}, vol.~96, no.~34, 1996, pp. 226--231.

\bibitem{distance_correlation}
G.~J. Sz{\'e}kely, M.~L. Rizzo, and N.~K. Bakirov, ``{Measuring and testing
  dependence by correlation of distances},'' \emph{Ann. Stat.}, vol.~35, no.~6,
  pp. 2769 -- 2794, 2007.

\bibitem{Mirjalili2019}
S.~Mirjalili, \emph{Genetic Algorithm}.\hskip 1em plus 0.5em minus 0.4em\relax
  Cham: Springer International Publishing, 2019, pp. 43--55.

\bibitem{gad2023pygad}
A.~F. Gad, ``Pygad: An intuitive genetic algorithm python library,''
  \emph{Multimedia Tools and Applications}, pp. 1--14, 2023.

\end{thebibliography}

\end{CJK}
\end{document}